\documentclass[lettersize,journal]{IEEEtran}
\usepackage{xcolor,soul,framed} 
\colorlet{shadecolor}{yellow}
\usepackage[pdftex]{graphicx}
\usepackage{float}
\graphicspath{{../pdf/}{../jpeg/}}
\DeclareGraphicsExtensions{.pdf,.jpeg,.png}
\usepackage[cmex10]{amsmath}
\usepackage{array}
\usepackage{mdwmath}
\usepackage{mdwtab}
\usepackage{eqparbox}
\usepackage{url}
\usepackage[ruled]{algorithm2e}
\usepackage{hyperref}
\hypersetup{hypertex=true,
            colorlinks=true,
            linkcolor=black,
            anchorcolor=black,
            citecolor=black}
\usepackage{cite}
\usepackage{amsthm,amsmath,amssymb}
\usepackage{mathrsfs}
\def\BibTeX{{\rm B\kern-.05em{\sc i\kern-.025em b}\kern-.08em
		T\kern-.1667em\lower.7ex\hbox{E}\kern-.125emX}}

\usepackage{multirow}
\usepackage{balance}
\usepackage{subfigure}

\begin{document}
\title{Exploring Task-oriented Communication in Multi-agent System: A Deep Reinforcement Learning Approach}
\author{Guojun He
	
\thanks{Guojun He is with the Research Center of 6G Mobile Communications, Wuhan National Laboratory for Optoelectronics, Huazhong University of Science and Technology, Wuhan 430074, P.R.China.}
}

%
\maketitle

\begin{abstract}
The multi-agent system (MAS) enables the sharing of capabilities among agents, such that collaborative tasks can be accomplished with high scalability and efficiency. MAS is increasingly widely applied in various fields. Meanwhile, the large-scale and time-sensitive data transmission between agents brings challenges to the communication system. The traditional wireless communication ignores the content of the data and its impact on the task execution at the receiver, which makes it difficult to guarantee the timeliness and relevance of the information. This limitation leads to that traditional wireless communication struggles to effectively support emerging multi-agent collaborative applications. Faced with this dilemma, task-oriented communication is a potential solution, which aims to transmit task-relevant information to improve task execution performance. However, multi-agent collaboration itself is a complex class of sequential decision problems. It is challenging to explore efficient information flow in this context. In this article, we use deep reinforcement learning (DRL) to explore task-oriented communication in MAS. We begin with a discussion on the application of DRL to task-oriented communication. We then envision a task-oriented communication architecture for MAS, and discuss the designs based on DRL. Finally, we discuss open problems for future research and conclude this article.
\end{abstract}

\IEEEpeerreviewmaketitle

\section{INTRODUCTION}
In recent years, with the rapid development of wireless communication, electronic and artificial intelligence (AI) technologies, multi-agent system (MAS) is increasingly applied to various areas, such as precision agriculture with drones, smart manufacturing, and autonomous driving. Different from the centralized control approaches, MAS decentralizes the decision-making ability from the central controller to the spatially distributed agents. Each agent is independent and has the ability to sense and execute. Meanwhile, the agents can interact with each other through communication, and collaborate to accomplish the task. Distributability and cooperation makes MAS more scalable and efficient for complex tasks.

Multi-agent collaboration relies on the sharing of behavioral intentions and environmental observations. The data that contains this information is generally multi-modal and time-sensitive. For example, in the autonomous driving scenario, each vehicle generates at least 750 MB of data per second. The generated data includes multi-modal observations of the environment such as video and radar point clouds, as well as observations of its own state such as position and velocity. The vehicles' states are time-sensitive and need to be sent to surrounding vehicles timely to avoid collisions. In addition, the communication between agents is not an end in itself but a means to foster multi-agent collaboration. Transmitting task-irrelevant information does not contribute to the collaboration, and even causes network congestion, wasting wireless resources\cite{9475174}. These characteristics impose new requirements on communication system design.

The traditional wireless communication system follows the data-oriented principle that pursues reliable and efficient data transmission. The traditional communication generally only focuses on the data transmission process, ignoring the generation and utilization of the data. Data significance is separate from data transmission. This nature makes it hard for traditional communication to deal with the coupling relationship between communication strategies and downstream collaborative tasks. As a consequence, traditional communication struggles to guarantee the data at the receiver is fresh and relevant to the end task. This limits the traditional communication to support large-scale machine intelligence-like communication and emerging multi-agent cooperative applications.

To break through this dilemma, the design principle of the communication system needs to be updated. The classical principle that “semantic aspects of communication should be considered as irrelevant to the engineering problem”, is not suitable for multi-agent communication\cite{9771334}. Recently, a new communication paradigm named semantic communication is emerging in the community. Semantic communication can extract and transmit semantic information, relying on advanced signal processing techniques empowered by AI. Related researches involve semantic-oriented and task-oriented principles. The semantic-oriented principle aims to transmit semantic information precisely. The task-oriented principle focuses more on the significance of data to the end task, and aims to enhance the performance of task execution at the receiver. The nature of task-centric, makes task-oriented communication a promising way to support emerging multi-agent collaboration applications, in the context that wireless resources are increasingly scarce. However, multi-agent collaboration itself is a complex class of sequential decision-making problems. What's more, various collaborative tasks have different constraints, such as action, communication resource, and computational resource constraints. In this context, it is non-trivial to analyze the impact of communication and explore the optimal communication strategy for multi-agent collaboration.

Deep reinforcement learning (DRL) is a new class of intelligent decision-making algorithms. It combines the self-learning ability of reinforcement learning (RL) and the approximation ability of deep learning (DL), thus can search for optimal action policy in high-dimensional action space without any prior knowledge. DRL can adapt to dynamic environments, and has registered tremendous success in solving some complex problems, such as spectrum access, transmission power control, and user association. As a result, DRL is increasingly applied in wireless communication. In this article, we exploit DRL to explore task-oriented communication in multi-agent cooperative scenarios, and consider the implementation of related signal processing, wireless resource management (WRM), and behavioral decision-making.

In the remainder of this article, we first introduce the task-oriented communication model, and review recent applications of DRL in the task-oriented communication field. Then we envision a task-oriented communication framework for the MAS, and discuss the designs based on DRL. Following that we present a case study. Finally, we discuss open research problems and conclude this article.

\section{DEEP REINFORCEMENT LEARNING FOR TASK-ORIENTED COMMUNICATION}
\subsection{Point-to-point Task-oriented Communication Model}
\begin{figure}[htbp]
	\centering
	\includegraphics[width=3.46in]{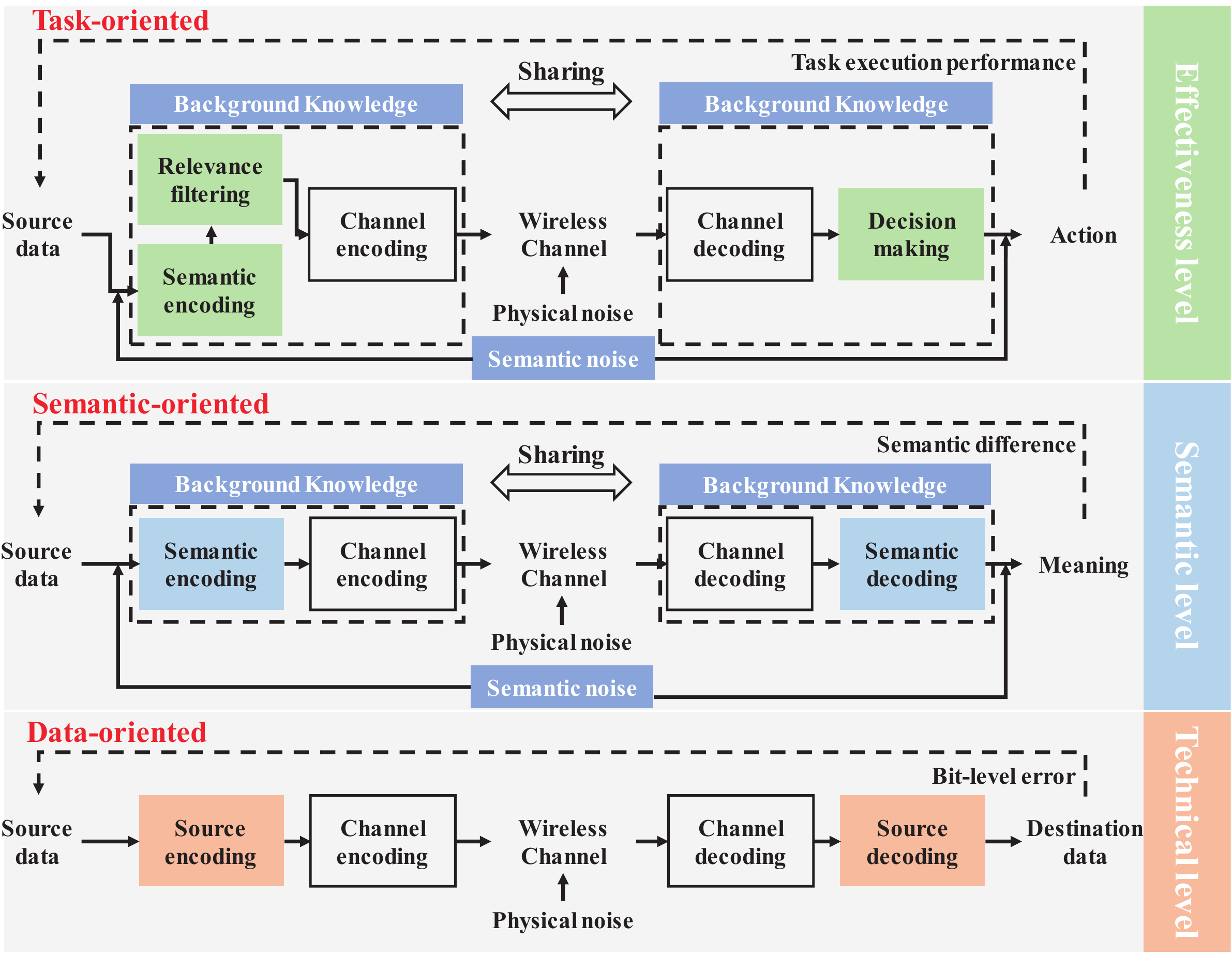}
	\caption{Point to point communication model with different design principles.}
	\label{fig_1}
\end{figure}

Shannon and Weaver classified communication problems into three levels: technical level, semantic level, and effectiveness level\cite{informationtheory}. According to their vision, task-oriented communication belongs to the effectiveness level. It origins from  data-oriented and semantic-oriented communication, and makes further extensions, that is from information transmission to information utilization. As shown in Fig.\ref{fig_1}, the core goal of the task-oriented principle is facilitating the sharing of task-relevant information, to improve the performance of task execution at the receiver.

In the point-to-point task-oriented communication model, the transmitter firstly performs semantic encoding to extract semantic information from the source data. Then the semantic information will be filtered for relevance, forming semantic features. On the one hand, this can reduce the occupation of communication resources and computing resources by task-irrelevant information. On the other hand, this can avoid potential interference with task execution from task-irrelevant information. As for the semantic features, they need to satisfy the following three conditions. Firstly, there is a consensus on the inference process of semantic features by the transceivers, that is, there is no ambiguity in the understanding of semantics. This is a prerequisite for semantic transmission, supported by the sharing of background knowledge between the transceivers. Secondly, the semantic features are able to represent the task-relevant semantics contained in the source data completely, avoiding the loss of information. Finally, semantic features are abstract and compact to save transmission costs. Note that, semantic information is interfered with by semantic noise during the transmission process, which would cause a misunderstanding of the semantic information by the receiver. Semantic noise can come from bit transmission errors caused by physical noise, or from differences in background knowledge between the transceivers. At the receiver, the decision maker directly uses decoded semantic features to make decisions. The recovery of semantic information is integrated into the decision-making process. 

The specific task goal plays an important role in the design of task-oriented communication systems. It determines the definition of task relevance and the specific way in which data is used. From this perspective, communication is no longer just at the front of task execution independently, but in the process of task execution.
\begin{figure*}[htbp]
	\centering
	\includegraphics[width=4.8in]{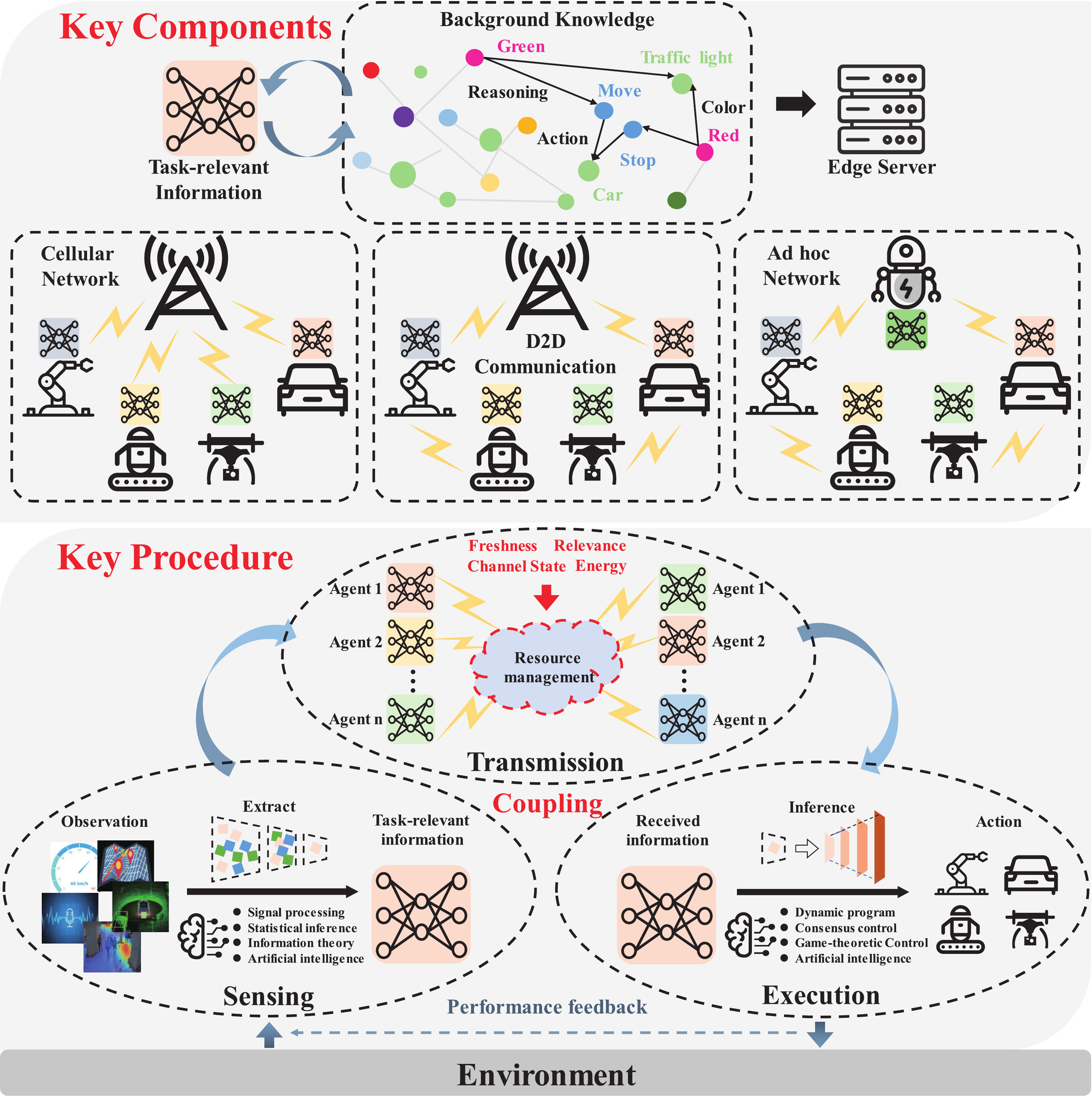}
	\caption{Task-oriented communication architecture for the MAS.}
	\label{fig_2}
\end{figure*}

\subsection{Technical Background of Deep Reinforcement Learning}
Generally, DRL models the sequential decision-making problem as a Markov decision process (MDP), which contains basic elements such as observations, actions, and rewards. At each time step, the agent observes the instantaneous environmental state, then selects an action using the current policy. The policy is constructed by DL technology. Specifically, in value-based algorithms, DRL uses deep neural networks (DNNs) to approximate the true value function. Relying on the estimated value function, the agent will choose the action with the highest value. In policy-based algorithms, DRL uses DNNs to approximate the optimal action policy directly, as a function of observation. After executing the selected action, the environment will transition to next state. Meanwhile, the agent receives a numerical reward that quantifies its contribution to the goal at this time step. The agent will gradually revise its action policy in a trial-and-error manner, to maximize the long-term cumulative reward.

DQN is one of the most effective value-based algorithms and is mainly applied to solve decision-making problems with discrete action spaces\cite{mnih2015human}. DQN uses DNNs to learn the state-action value function. To stabilize the learning process, DQN introduces the experience replay mechanism (ERM) and target Q-value. The ERM can eliminate the correlation between training data, and the target Q-value can slow down the update of model parameters. When sampling, the experience is stored into a buffer. When training, a batch of samples will be randomly sampled from the buffer. With these samples, DQN updates the model by minimizing the gap between the estimated state-action value and the target Q-value. DDPG belongs to policy-based algorithms\cite{lillicrap2015continuous}. DDPG introduces the actor-critic (AC) framework. Actor and critic are both composed of DNNs, where the actor is used to approximate the optimal action policy, and the critic is used to evaluate the performance of actor. Since the actor establishes a mapping from observation space to action space, DDPG is suitable for solving decision-making problems with continuous action spaces. The parameter update of actor is based on the deterministic policy gradient theorem, and the parameter update of the critic is similar to the way of DQN.

Multi-agent deep reinforcement learning (MADRL) as the extension of DRL, holds considerable promise to address the sequential decision-making problem of multiple agents that operate in a common environment. MADDPG is an extension of DDPG\cite{lowe2017multi}. To solve the challenge of non-stationarity in the multi-agent domains, MADDPG introduces the mechanism of centralized training with distributed execution (CTDE) based on the AC framework. During the training phase, the centralized critic evaluates the performance of each agent's policy based on global environmental states, guiding their updates. In the execution phase, each agent selects actions based on its local observations only. QMIX is an extension of DQN, which also follows the CTDE mechanism\cite{rashid2018qmix}. To efficiently extract decentralized policies from the centralized policy, QMIX enforces joint action value to be monotonic in the action value of each agent. This setting ensures that the optimal centralized policy is equivalent to the collection of per-agent optimal decentralized policy. This consistency can solve the problem of credit assignment among multiple agents.

\subsection{DRL for Task-oriented Communication}
Task-oriented communication has enormous potential, however, is challenging to be implemented. Specifically, to date the semantic information theory is incomplete. The construction of semantic information extractors and filters lacks guidance. In addition, task-oriented communication takes the task completion degree as the core. This leads to the optimal message generation and transmission strategies changing with task goals and environment. Thus, it is complex to handcraft communication strategies. On the other hand, the metric of message significance is also a critical concern in task-oriented communication. The metrics can be differentiable. However, non-differentiable ones are more common in practical wireless scenarios. Non-differentiable objectives are difficult to be optimized in existing task-oriented communication approaches that are based on end-to-end architecture. To solve these issues, DRL is a promising solution. DRL is model-free and can acquire knowledge by interacting with the dynamic environment. This enables DRL to explore the optimal communication strategy autonomously. Additionally, DRL is reward-driven. DRL can optimize the non-differentiable objectives by integrating them into the reward.

Due to this promising prospect, DRL has recently been applied to task-oriented communication systems. In \cite{9796572}, the authors consider a scenario with UAV-assisted (unmanned aerial vehicle) edge intelligence. The UAV collects images and sends them back to an edge server for intelligent analysis. To support the efficient analysis, the communication between the UAV and the edge server needs to be ultra-reliable and low-latency. To this end, the authors utilize DRL to construct a task-oriented transmission strategy. The UAV identifies and transmits the part of the original image that is most conducive to the analysis task, according to the image content as well as the channel gain. In \cite{9771334}, the authors introduce DRL for sentence transmission. Instead of the bit-level learning objectives such as mean-square error, the authors use semantic-level metrics such as BLEU to measure the sentence semantic similarity. These semantic-level metrics can more accurately reflect the semantic transmission performance, however, are non-differentiable. Thus, they can not be optimized by gradient descent algorithms directly, as in supervised learning. To this end, the authors use the metrics to design rewards and accomplish end-to-end learning of the communication system with the AC framework.

The above works are oriented to non-sequential decision-making problems, and consider task-oriented communication from the perspective of wireless communication. For sequential decision-making problems in multi-agent collaboration, there are many excellent recent works exploring the effectiveness of communication from the perspective of computer science. In \cite{NIPS2016_c7635bfd}, the authors consider fully collaborative, partially observable, multi-agent sequential decision problems. Each agent needs to share with others the information needed for collaboration. To develop an efficient and effective communication protocol between the agents, the authors try for the first time a self-learning approach. The agents need to learn how to make action decisions and what messages to send. To achieve this, the authors adopt the DQN and train it based on CTDE. In \cite{kim2019learning}, the authors explore the scheduling problem in multi-agent communication. Multiple agents need to communicate over a shared wireless channel during the collaboration process. To solve the problem of potential access conflicts, the authors construct a communication scheduling framework named SchedNet based on MADRL. SchedNet selects a subset of agents based on the importance of their observed information and allows them to broadcast messages simultaneously. In \cite{9466501}, the authors consider the problem of reducing the interference of channel noise on the effectiveness of multi-agent communication. The authors design a joint source-channel coding scheme, and test it on the BSC channel, AWGN channel and BN channel. The authors verify that the performance of multi-agent collaboration when jointly optimizing communication and decision-making is superior to that when considering communication separately from decision-making. 

\section{TASK-ORIENTED COMMUNICATION FOR MULTI-AGENT SYSTEM}
A MAS consists of multiple agents with sensing, communication, and computing capabilities. The agents can complement and share capabilities through cooperation, and thus have higher efficiency and robustness when facing complex tasks, than the single-agent system. Collaboration relies on communication. The agents need to share observational information to revise their cognition of the environment, avoiding making sub-optimal decisions. They also need to share behavioral intentions for efficient negotiation to avoid behavioral conflicts. These information interactions are required to be timely and task-relevant. However, the data-oriented communication system is blind to the content and timeliness of the message. It cannot judge whether the transmitted data is relevant to the task. The existing KPIs, such as bit error rate and latency, cannot measure the freshness of the data. As a consequence, in the MAS with the data-oriented principle, there can be a large amount of outdated, task-irrelevant data at the receiver. These data are helpless for task execution and occupy increasingly scarce wireless resources, causing communication bottlenecks. To this end, we propose a task-oriented communication architecture for the MAS, as shown in Fig.\ref{fig_2}.

\subsection{Task-oriented Communication Architecture}
\subsubsection{Key Components}
In the proposed architecture, the agents are actual executors. They can be homogeneous or heterogeneous, such as unmanned vehicles and UAVs. The agents are independent and autonomous, meanwhile interacting with each other implicitly. They operate in a common environment and are expected to accomplish the desired collaborative tasks efficiently, with limited time, communication resources, and computational resources.

The agents exchange information through wireless communication. According to the specific task scenario, data transmission can be realized via cellular communication, D2D communication, ad hoc network, etc. For example, in the UAV-assisted edge intelligence scenario, the UAV can use the cellular network to transmit data back to the back-end server  \cite{9796572}. In the autonomous driving scenario, the unmanned vehicle can use cellular links to transmit large volumes of multimedia data, while using V2V links to transmit time-sensitive data such as location and speed. In the rescue scenario, infrastructure such as base stations is not available. UAVs can form an ad hoc network to transmit messages.

In the interaction of agents, task-relevant information is presented in the form of semantic features. The premise that semantic features can be transmitted is that there is no ambiguity in the understanding of semantics among the agents. For example, in the autonomous driving scenario, when the traffic signal is observed to be red, the unmanned vehicle will stop; when the traffic signal is observed to be green, the unmanned vehicle will keep moving. This consensus relies on the existence of an intersection among agents' background knowledge. The background knowledge can be presented in the form of a knowledge graph. It contains real-world entities, relationships between entities, possible ways of understanding and reasoning, and so on. Each agent can continuously learn and update the background knowledge. In addition, when the computing power and storage capacity are insufficient, the agents can offload the background knowledge to the edge server. When there is a large difference between the background knowledge, the agents can also negotiate through the edge server to update them.  

\subsubsection{Key Procedure}
In the proposed architecture, the process of multi-agent collaboration can be mainly divided into three parts: sense, transmission, and execution. The three parts influence each other and are coupled in a closed loop. The agents will automate decisions in the cycle of the three. Among them, the principle of task-oriented communication is mainly used to answer three basic questions, what messages each agent needs to send, when to send them, and to whom.

To be specific, in the sensing phase, each agent uses multiple types of sensors to observe the surrounding environment as well as its current state, and then records the observed data such as video, radar point cloud, and depth map. Due to limited sensor capabilities and practical factors such as object occlusion, these observation data generally can only capture part of the environmental state. The partial observability of the environmental state drives the agents to communicate. The expectation is to develop a more comprehensive cognition of the environment to facilitate decision-making. However, multi-modal observation data are large in volume. It is not practical to transmit all the raw data directly. Therefore, to efficiently transmit and utilize the observation data, each agent will perform multi-modal fusion to them. The data with different modalities will be mapped into the semantic space supported by the same background knowledge, which in turn filters out the task-relevant semantic information to form semantic features.  

In the transmission phase, the agents transmit the semantic features by wireless communication. Generally, wireless resources are scarce, and wireless signal transmission conditions are unstable due to factors such as fading and interference. In addition, in some scenarios, it is also necessary to minimize the interference of multi-agent communication to other users, such as in cognitive radio networks. Therefore, the wireless communication system needs to use advanced spectrum allocation, power control, access control, scheduling, and other techniques to perform efficient resource management to ensure efficient and reliable transmission of semantic features. Different from the traditional WRM, under the task-oriented principle, the resource management strategy is associated with the generation and usage of messages. The communication system needs to comprehensively consider the relevance, freshness, channel state, energy consumption, and other factors to determine the priority of message transmission. And then, the flow of information is reasonably regulated to realize that semantic features are transmitted to the agents that need it at the right time, laying the foundation for improving the performance of multi-agent collaboration.  

In the execution phase, each agent extracts semantic information from the received semantic features. Based on this semantic information, each agent will infer the whole picture of the task environment, as well as the state and intention of other agents, and then make the behavioral decision. In addition, multi-agent collaboration is a sequential decision-making problem. The behavioral decisions at each time step manipulate the environment, which will provide performance feedback to the MAS and influence the next round of the Sensing-Transmission-Execution cycle.

\subsection{Design Approaches with DRL}
\subsubsection{Block Design}
In the proposed architecture, accurate measurement of data significance is essential for achieving task-oriented communication. Recently, the concept of age of information (AoI) has received a lot of attention. It is defined as the time that has elapsed since the last received packet at the receiver was generated at the transmitter. Different from the metric of delay, AoI can quantify the freshness of data. In some multi-agent cooperative applications, such as autonomous driving and UAV swarm control, where the freshness of data has a significant impact on the decision-making of agents, AoI can also be utilized to measure the effectiveness of data. From this perspective, the MAS can construct the communication strategy with the goal to minimize AoI. The data will be driven to transmit to the right agents at the right time, and then improving the collaborative performance of the MAS. As for the complex decision issues in WRM, they can be solved with help of DRL. Taking the flocking task as an example, the authors in \cite{9636873} use graph neural networks (GNNs) to construct a data distribution strategy for ad hoc communication. Each agent can choose which agent to communicate with, based on the local data structure. In addition, the authors exploit AoI to construct the reward and train the model with the CTDE architecture. Finally, the flocking can be performed better with the proposed data distribution strategy, than that with data-oriented policies such as random flooding and round robin.

In addition to AoI, there are many researches, such as age of incorrect information and value of information, have explored how to measure the meaning of data from different perspectives \cite{uysal2021semantic}. Built on these handcrafted metrics, task-oriented communication is generally concise and flexible. Because it can be separated from the decision-making for back-end tasks in terms of design and optimization. As for the back-end tasks, they can be solved by the currently available theories and algorithms such as dynamic planning and consensus control, according to specific application scenarios.

\subsubsection{End-to-End Design}
The simplicity and flexibility of the block design are at the expense of final performance. Another design, which can improve performance, is the end-to-end design based on MADRL architecture. In the end-to-end design, task-oriented communication is no longer a part to be designed and optimized separately, but is embedded in the MADRL architecture. The generator of semantic features, the scheduler of wireless resources, and the decision maker of task execution are all built with DNNs. Note that, the building process needs to consider practical constraints such as action space, communication and computation resources. In addition, these constraints will be combined with specific task performance evaluation metrics, such as execution time and success rate, to design rewards. By maximizing the rewards, the optimal collaborative performance of the MAS with the given resources is achieved. As for the model training, CTDE is a practical architecture.

End-to-End Design is dedicated to a specific task. While achieving high performance, it also loses the universality. In addition, the joint optimization of communication and action policy for task execution poses a considerable challenge to the algorithm design of MADRL. 

\subsection{Case Study}
As a case study, in this subsection, we present how to implement end-to-end design for task-oriented communication in a multi-agent cooperative task.
 
We consider a networked MAS, which consists of one access point (AP) and four agents. The AP is assumed to have two orthogonal subchannels to support the uplink transmissions of the agents. As for the downlink transmission, the AP will broadcast the messages. The channel model is same as \cite{9847027}. The agents are required to perform the Predator and Prey\cite{9597491}, that is chasing four random walks preys within a certain area with obstacles. Here, we simplify the environment to a $20 \times 20$ grid world. The agents have five action choices per time step: up, down, left, right, or stay. In addition, each agent has a $7 \times 7$ field of view, in which the agents can observe the position of other agents, prey, and obstacles. The whole system operates in a sensing-transmission-execution cycle. To efficiently control the flow of information, here we adopt our proposed model TOCF\cite{9847027}. On the transmitter side, each agent generates semantic features following the information bottleneck principle, i.e., extracting relevant information and discarding redundant information. For channel allocation, the AP is also considered as an agent, that chooses allocation actions based on the real-time channel states and the importance of semantic features. On the receiver side, each agent differentially fuses the received semantic features, and then makes action decisions. Here, we design the rewards relying on task execution performance, such as the capture of prey and message transmission time.
\begin{figure}[htbp]
	\centering
	\subfigure[Success Rate]{
		\begin{minipage}[htbp]{0.48\linewidth}
			\centering
			\includegraphics[width=1.75in]{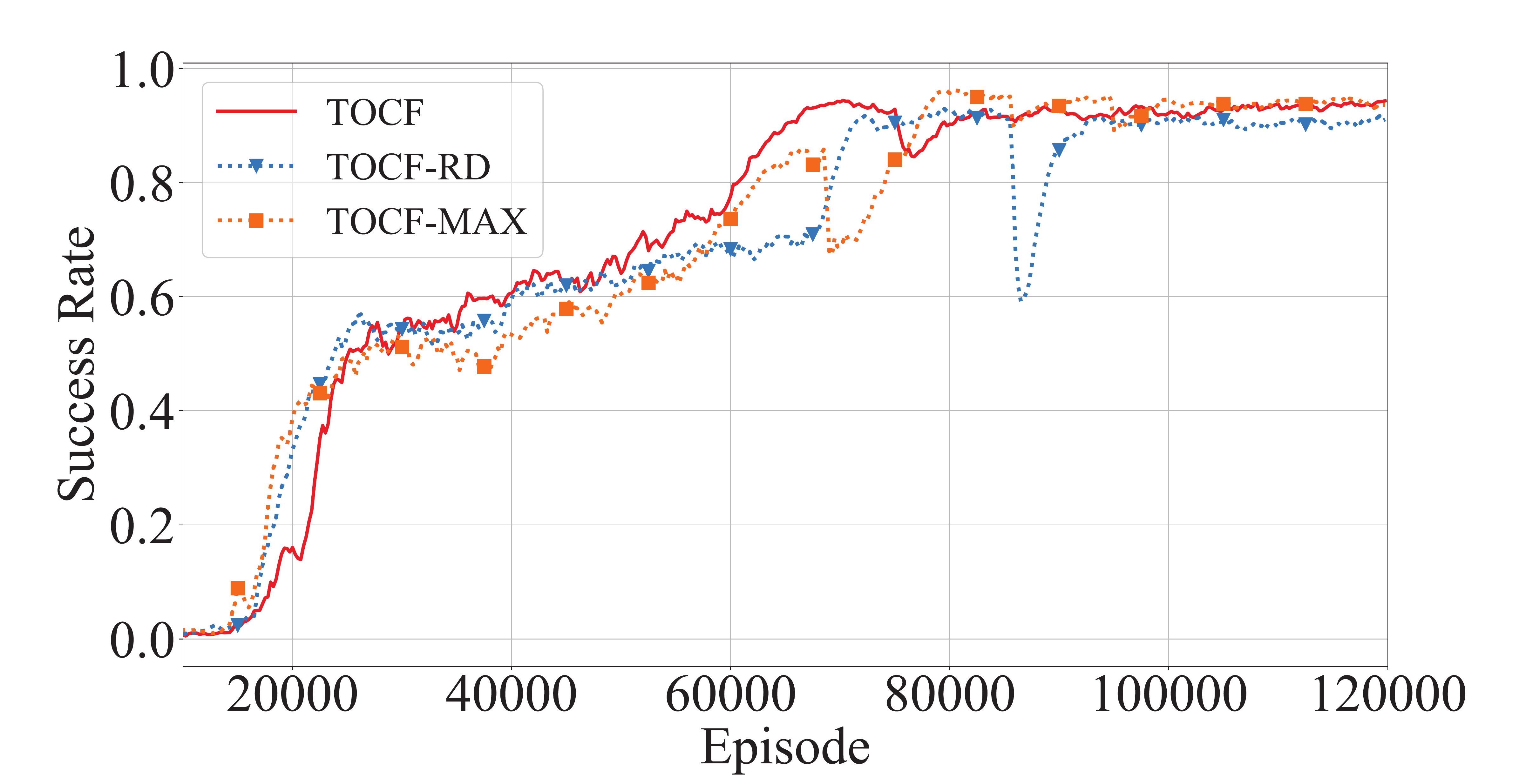}
		\end{minipage}%
	}%
	\subfigure[Episode Total Time]{
		\begin{minipage}[htbp]{0.48\linewidth}
			\centering
			\includegraphics[width=1.75in]{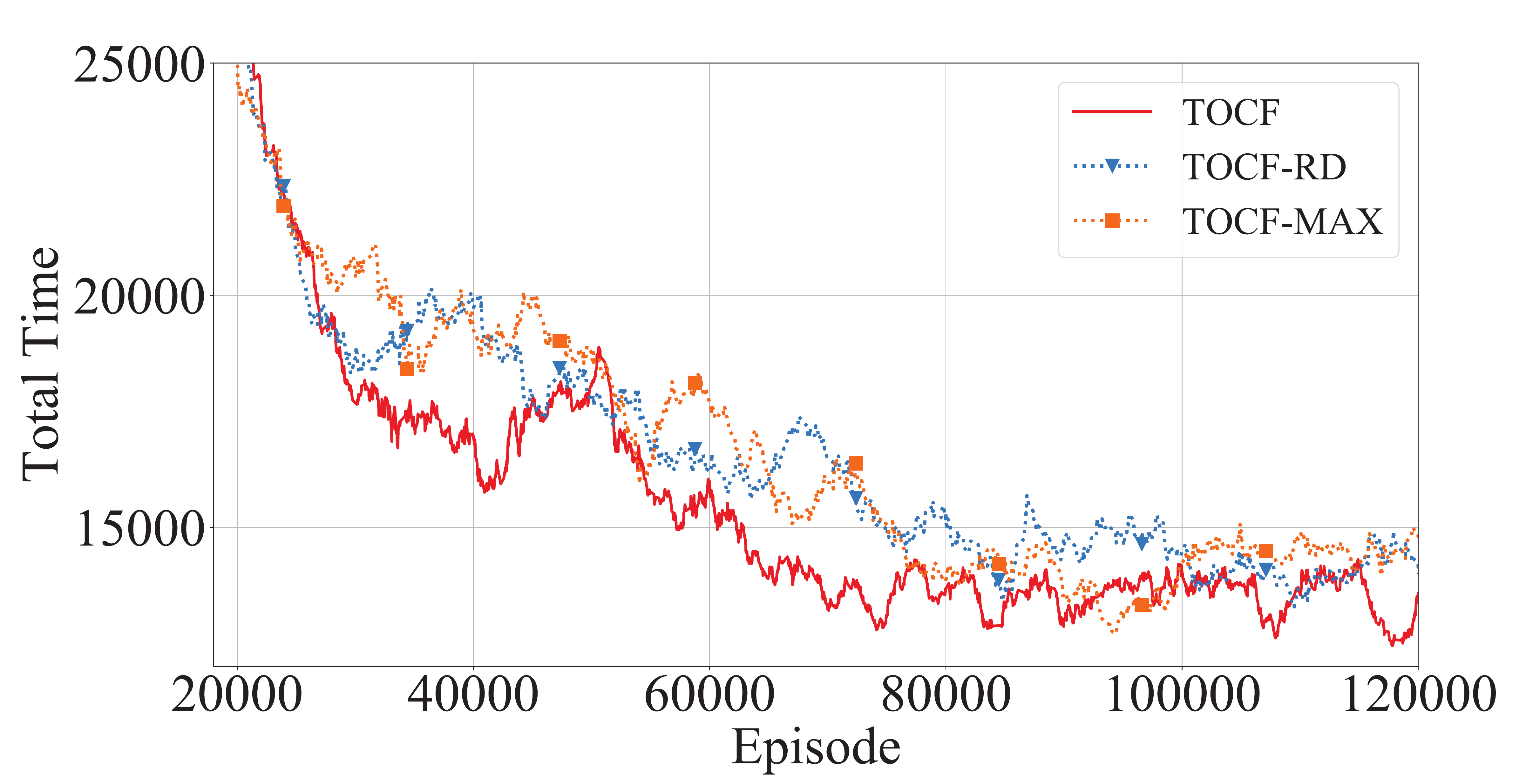}
		\end{minipage}%
	}%
	\caption{Learning curves with respect to different wireless channel allocation methods.}
	\label{fig_3}
\end{figure}

\begin{figure}[htbp]
	\centering
	\includegraphics[width=3in]{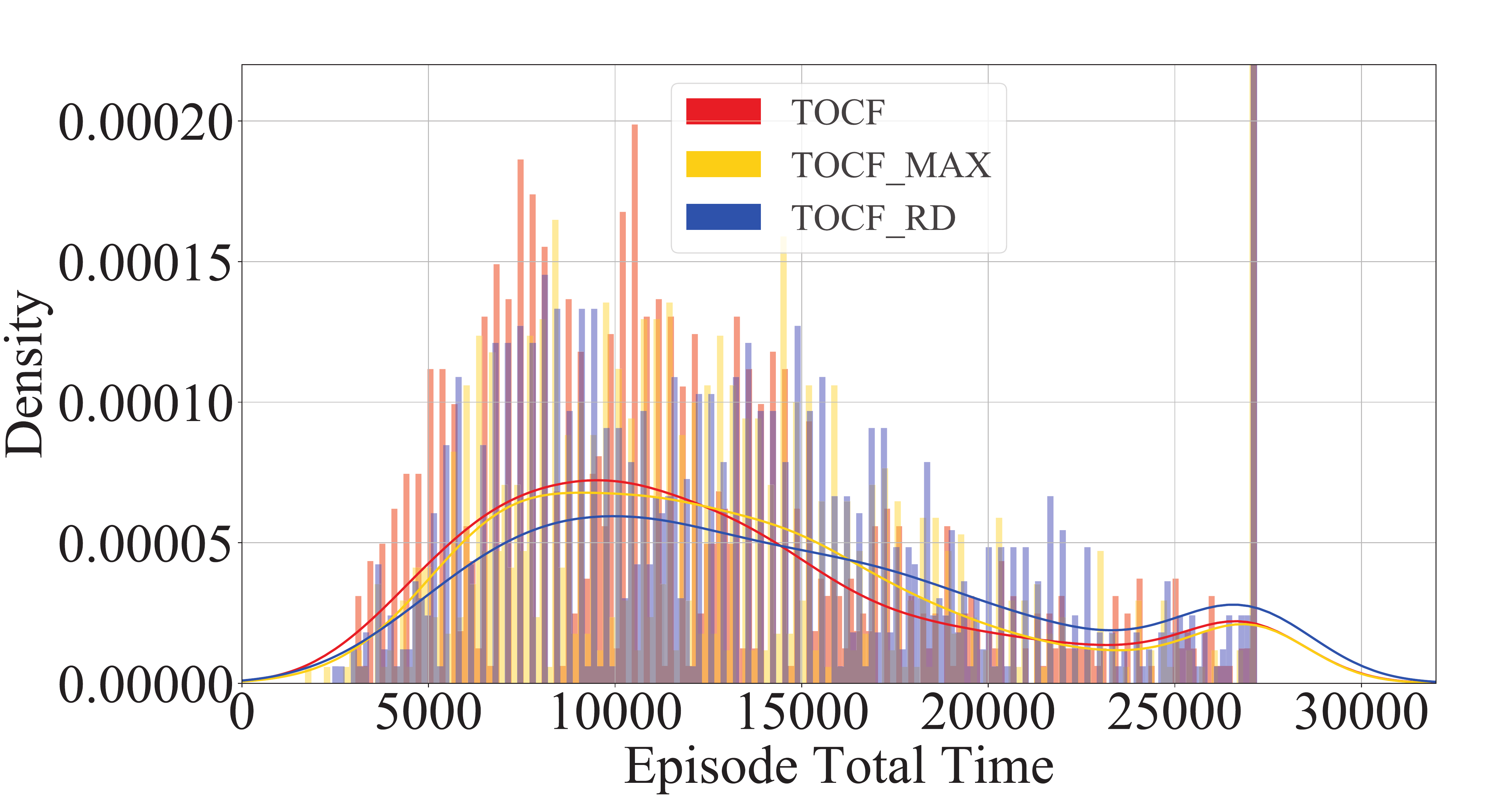}
	\caption{Density plots of episode total time with respect to different wireless channel allocation methods.}
	\label{fig_4}
\end{figure}

The Fig.\ref{fig_3}. shows the learning curves with respect to different wireless channel allocation methods. In the TOCF-RD, the AP allocates channels randomly. In the TOCF-MAX, the AP allocates channels with the goal to maximize the message transmission rate. We train the three methods in the environment with regularly arranged and fixed density obstacles. We can see that, the three methods start to stabilize at around 80000 episodes, and finally, the TOCF has the lowest task execution time. The Fig.\ref{fig_4}. shows the density plots of episode total time with respect to different wireless channel allocation methods. We test the three methods in the environment with dynamic density obstacles over 1000 episodes. We can see that the TOCF has the lowest mean episode total time, showing that TOCF has the best ability to adapt to the dynamic environment. These two tests show the effectiveness of task-oriented channel assignment. Joint optimization of communication and action policies can better regulate the flow of task-relevant information, and then improve the effectiveness of communication.

\section{OPEN PROBLEMS AND FUTURE RESEARCH}
\textbf{Complete information theory:} The current theoretical basis for task-oriented communication is incomplete. The classical information theory exploits uncertainty to measure the information amount. However, this is not sufficient to accurately measure the semantic information amount of the data, and its relevance to the end task. In this regard, it is necessary to break through the limitations of classical information theory, and establish semantic and pragmatic information theory. Defining the measurement form of data significance, based on the attributes of data, to provide theoretical guidelines for the extraction of semantic information as well as the representation of semantic features. In addition, it is also necessary to establish a measurement system for network performance, in which studying the trade-off relationship between data significance, channel state, available wireless resources, and agent on-board resources to provide theoretical guidance for efficient WRM.

\textbf{Neural network with strong representation capability:} Both in block design and end-to-end design, the underlying neural network structure is an essential component. It largely determines the performance of the whole system. Most of the existing work for MADRL, uses fully connected multi-layer perceptrons (MLPs) or convolutional neural networks (CNNs). MLPs and CNNs have shown excellent performance in processing Euclidean data, but they have shortcomings when facing non-Euclidean data, such as wireless network topology. They cannot capture the spatial, logical relationships between the communicating nodes well. In this regard, GNNs are potential solutions. GNNs learn the representation of graph data through message passing between nodes, and can take into account both edge features and node features. In addition, GNNs are able to handle graphs of different edges and nodes, and then can adapt well to dynamically changing wireless network topology.

\textbf{Efficient training mechanism:} The essence of RL is trial and error. The agents need more samples to improve their actions, in the case of large action space and state space. This imposes a challenge for many practical cooperative applications. In this regard, it is necessary to construct efficient sampling strategies to reduce the amount of data required for model training. In addition, imitation learning can provide prior knowledge to the agents, which is a potential solution. On the other hand, in real scenarios, the agents are generally spatially distributed. Thus, the deployment of the CTDE mechanism inevitably requires information interaction. At this time, in addition to the pursuit of high communication efficiency, the security and privacy of data transmission also need to be guaranteed. In this aspect, federated learning technology can play a role.

\section{CONCLUSION}
In this article, we explore task-oriented communication in MAS with DRL. We first discuss the background of task-oriented communication, and review the applications of DRL in the task-oriented communication field. Then we propose a task-oriented communication architecture for MAS. We also discuss the designs and show the effectiveness through a case study. Finally, we discuss the future research problems and conclude this article.

\ifCLASSOPTIONcaptionsoff
\newpage
\fi

\bibliographystyle{IEEEtran}
\bibliography{IEEEabrv,ref}
\vfill

\section*{Biographies}
\textbf{Guojun He (guojunhe@hust.edu.cn)} is currently pursuing his Ph.D. degree with the Research Center of 6G Mobile Communications, Wuhan National Laboratory for Optoelectronics, Huazhong University of Science and Technology, Wuhan, P.R.China. His research interests include wireless communication and deep reinforcement learning.

\end{document}